\documentclass[twocolumn,nopacs,preprintnumbers,amsmath,amssymb]{revtex4}

\usepackage{verbatim}
\usepackage{graphicx}
\usepackage{dcolumn}
\usepackage{bm}

\begin{document}

\title{Tungsten oxide nanowire growth by chemically-induced strain}
\author{Christian Klinke}
\email{klinke@chemie.uni-hamburg.de}
\author{James B. Hannon}
\author{Lynne Gignac}
\author{Kathleen Reuter}
\author{Phaedon Avouris}
\affiliation{IBM T. J. Watson Research Center, 1101 Kitchawan Road, Yorktown Heights, NY 10598, USA}

\begin{abstract}

We have investigated the formation of tungsten oxide nanowires under different CVD conditions. We find that exposure of oxidized tungsten films to hydrogen and methane at 900$^{\circ}$C leads to the formation of a dense array of typically 10~nm diameter nanowires. Structural and chemical analysis shows that the wires are crystalline WO$_3$. We propose a chemically-driven whisker growth mechanism in which interfacial strain associated with the formation of tungsten carbide stimulates nanowire growth. This might be a general concept, applicable also in other nanowire systems.

\end{abstract}

\maketitle

Interest in semiconducting nanowires continues to grow, in part,
because of their potential in nanoelectronics and
optoelectronics~\cite{WANG}. To date, a wide variety of
nanowire-based devices have been demonstrated, including
photodetectors, photodiodes, and sensors~\cite{VAYSSIERES}.
Nanowires offer certain advantages over planar devices. For
example, strain arising from lattice mismatch can be accommodated
via radial relaxation, rather than through the creation of bulk
dislocations, allowing new types of heterojuctions to be
formed~\cite{SAMUELSON}. The high surface/volume ratio can also
make nanowires efficient sensors~\cite{LIN}.

The majority of nanowires are grown using the vapor-liquid-solid
(VLS) mechanism, in which the nanowire grows from a small,
liquified gold catalyst particle~\cite{WHISKER1,WHISKER2}. The
size of the catalyst particle determines the diameter of the
nanowire. Tungsten oxide nanowires, on the other hand, grow by a
completely different mechanism that does not involve catalyst
particles. Several groups have shown that oxide whiskers can form
when tungsten substrates are heated in various mixtures of argon,
hydrogen, and oxygen~\cite{OKUYAMA,LI,LIU,LI2}. While it is clear
that the growth mechanism is not related to VLS, it is unclear how
and why the nanowires form. We have grown tungsten oxide nanowires
with a high-yield process that gives new insight into the driving
forces for nanowire formation. We find that exposure of oxidized
tungsten surfaces to first hydrogen and then methane increases the
yield of the nanowires by about two orders of magnitude compared
to a process that involves hydrogen, methane, or argon alone.
Chemical characterization shows that the wires are crystalline
WO$_3$ and that WC has formed on the substrate. We propose that
the formation of tungsten carbide at the surface strains the
surface oxide leading to enhanced whisker growth following a
mechanism described by Fisher \emph{et al.}~\cite{FISHER} and
Franks~\cite{FRANKS} in the 1950s. Our results show for the first
time that nanowire growth can be significantly enhanced by
chemically induced strain, in this case through the formation of
tungsten carbide at the substrate.

Tungsten oxide has received considerable attention due to its
application as a sensor material. Its bandgap lies in the optical
spectrum, making it attractive for optics applications. Tungsten
oxide has also been widely studied due to its `chromogenic'
properties~\cite{BANGE}. That is, the optical properties can be
varied by exposure to heat, light, or an applied potential. The
same properties make tungsten oxide a promising material in
applications ranging from sensors~\cite{PASSACANTANDO}, to ``smart
windows''~\cite{BERTRAN}.

\begin{figure} 
\begin{center}
\includegraphics[width=0.45\textwidth]{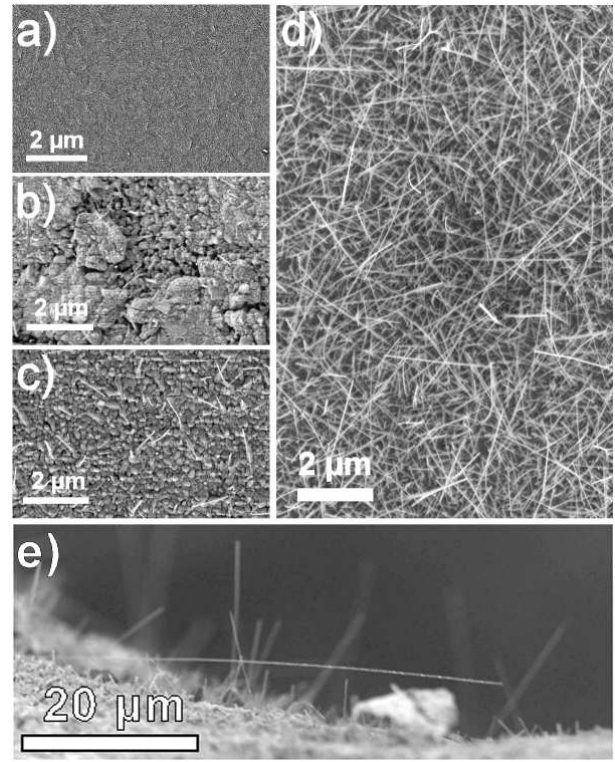}
\caption{\it SEM micrographs of a) the pristine tungsten
substrate, and nanowires grown at 900$^{\circ}$C with b) Ar, c) Ar
+ H$_{2}$, and d) CH$_{4}$ + H$_{2}$ in the CVD system. e) The
nanowires can be up to 40~$\mu$m long.} \label{F-CVD}
\end{center}
\end{figure}

We used two different CVD systems to investigate nanowire growth.
In the first, a standard 10~cm diameter tube furnace was used. The
substrates -- either tungsten wires, foils, or evaporated thin
films (100~nm tungsten on silicon) -- were introduced into the
furnace on a quartz holder. The furnace was evacuated to a base
pressure of 1~Torr. Our three-step process involved a heating
phase followed by a growth phase, and then slow cooling to room
temperature. In the heating phase, the temperature is ramped to
900$^{\circ}$C at a rate of 50$^{\circ}$/min in a gas flow of
500~sccm Ar and 200~sccm H$_{2}$ (total pressure of 100~Torr). In
the growth phase the temperature is held at 900$^{\circ}$C in a
mixture of 200~sccm hydrogen and 500~sccm methane for 30~min at a
pressure of 100~Torr. Finally, the samples were cooled to room
temperature in a gas flow of 500~sccm Ar.

\begin{figure} 
\begin{center}
\includegraphics[width=0.45\textwidth]{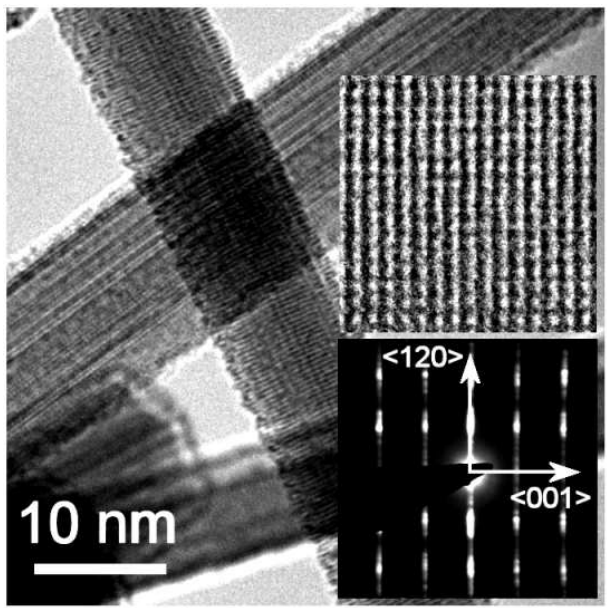}
\caption{\it TEM micrographs showing the lattice fringes and the
diffraction pattern (insets) of individual tungsten oxide
nanowires. The long axis of the wires points in $<$001$>$
direction of WO$_{3}$.} \label{F-TEM}
\end{center}
\end{figure}

In the first experiments, nanowires were grown by heating W films
(Fig.~\ref{F-CVD}a) in an Ar flow in the tube furnace. A scanning
electron microscopy (SEM) image of a film exposed to 500~sccm Ar
at 900$^{\circ}$C is shown in Fig.~\ref{F-CVD}b. Under these
growth conditions the film partially dewets and roughens. The
surface features are rounded and smooth (i.e.\ no facets), and
very few nanowires are observed. However, we found that by adding
H$_{2}$ to the gas flow (e.g. 500~sccm Ar + 200~sccm H$_{2}$), we
obtained a more uniform surface consisting of crystallites that
were clearly faceted (Fig.~\ref{F-CVD}c), while the nanowire yield
remained relatively low. A dramatic change in the growth occurs
when the growth stage flow includes methane (Fig.~\ref{F-CVD}d).
The yield of nanowires for the mixed flow is about two orders of
magnitude higher than when either hydrogen or methane are used
separately. The nanowires are often several microns long, with a
typical diameter of about 10~nm. Occasionally, we observed
extremely long nanowires of up to 40~$\mu$m in length
(Fig.~\ref{F-CVD}e).

We systematically varied the process conditions to investigate the
influence of temperature, gas pressure, growth time, and gas
mixture on the yield and length of the nanowires. We found that
nanowires grew in the temperature range from 600$^{\circ}$C to
1000$^{\circ}$C. However, only at 900$^{\circ}$C did we observe a
high yield, with wires longer than 1~$\mu$m. Above 900$^{\circ}$C
tungsten oxide starts to sublimate in vacuum~\cite{BANGE}.
Surprisingly, extending the duration of the growth phase did not
increase either the areal density or length of the wires.
Furthermore, the yield was essentially unchanged when the total
pressure of the growth phase mixture was varied from 10 to
500~Torr. Varying the methane/hydrogen ratio over a wide range
also had no significant influence on the wire length or diameter.
To summarize, high yield requires hydrogen pretreatment followed
by exposure to methane.

In order to determine the chemical composition and structure of
individual nanowires we grew them directly on transmission
electron microscopy (TEM) grids made of tungsten. TEM images show
that the wires are crystalline, with fringes visible both
perpendicular and parallel to the wire axis (Fig.~\ref{F-TEM}).
The corresponding lattice distances are 3.3~\AA~ and 3.7~\AA,
consistent with the structure of WO$_3$. The corresponding
diffraction image (Fig.~\ref{F-TEM} inset) suggests that the wires
are monoclinic WO$_{3}$, with the $<$001$>$ crystal axis oriented
parallel to the long axis of the wire. The streaking in the
direction orthogonal to the long axis indicates significant
twinning. Energy dispersive X-ray (EDX) measurements and electron
energy loss spectroscopy (EELS) performed directly on individual
nanowires confirmed the presence of only tungsten and oxygen in
the wires. X-ray photoelectron spectroscopy (XPS) was used to
investigate the chemical composition of the nanowires and
substrates after growth. The chemical shifts of the W 4f levels
are compatible with a mixture of WC and WO$_3$. Samples measured
before growth show no carbide peak. As we describe below, tungsten
carbide is formed on the W substrate during the exposure to
methane.

\begin{figure} 
\begin{center}
\includegraphics[width=0.45\textwidth]{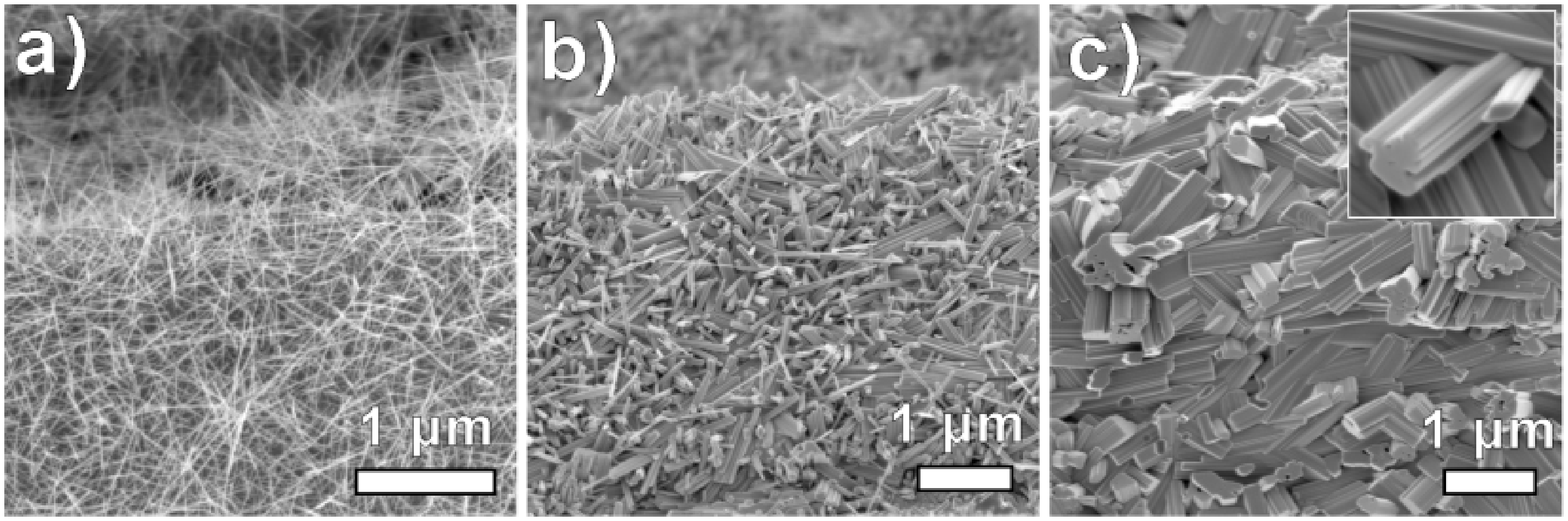}
\caption{\it Evolution of the morphology of the tungsten oxide
nanowire film grown in the vacuum system with increasing
temperature (from left to right). The comparison of pyrometer
measurements along the heated filament with the corresponding SEM
images allows to estimate the temperature in this series to be
between 800 and 950$^{\circ}$C.} \label{F-Temperature}
\end{center}
\end{figure}

In order to determine the source of the oxygen in the nanowires,
we grew them in a vacuum system with a base pressure of about
10$^{-7}$ Torr. The substrates consisted of resistively-heated W
filaments. Nanowires were grown by first backfilling the chamber
to 200~mTorr with H$_{2}$. The filament was then heated for 30~min
at 1000$^{\circ}$C (at the hottest point of the wire).
Subsequently, the chamber was pumped to 10$^{-7}$ Torr again and
CH$_{4}$ was introduced until the pressure reached 5~Torr. The
filament was heated for 30~min in the methane atmosphere, and then
cooled to room temperature. This process resulted in nanowire
growth that was qualitatively similar to that achieved in the tube
furnace: a high density of nanowires with a typical diameter of
about 10~nm (Fig.~\ref{F-Temperature}a). However, if the filament
was flashed to 2000$^{\circ}$C just prior to processing, no
nanowires were observed. We propose that the oxygen required for
nanowire formation comes from the native oxide on the W
substrates. The fact that the nanowire length can not be extended
by extending the growth time supports this conclusion.

In the vacuum system the filament was clamped to thick Cu
feedthroughs, resulting in a significant temperature gradient
along the length of filament. The gradient allowed us to
systematically investigate the role of temperature on the nanowire
growth under otherwise identical process conditions. A sequence of
SEM images along the filament length (i.e. with increasing
temperature) is shown in Fig.~\ref{F-Temperature}. As the
temperature is increased, the density and length of the nanowires
\emph{decreases}, while their diameter \emph{increases}. As we
describe below, this behavior is consistent with conventional
whisker growth.  Whiskers are non-equilibrium structures that grow
as a result of bulk dislocation motion (i.e. climb). In one
scenario described by Frank~\cite{FRANK}, the intersection of a
dislocation with the surface traces out a periodic loop.  Each
time the loop is traversed, the area of the surface enclosed by
the loop will be displaced by a Burger's vector.  This mechanims
gives rise to whiskers with irregular cross-sections but smooth
side walls, as shown in in Fig.~\ref{F-Temperature}c(inset). The
shape cross section is determined by the loop that the dislocation
core traces at the surface (the path is ultimately determined by
irregularities in surface morphology). Whisker growth is thought
to be governed by bulk vacancy motion (in this case, presumably
the tungsten oxide given the low mobility of vacancies in W at
1000$^{\circ}$C). In simple geometries~\cite{FRANK}, the whisker
growth rate is (approximately) proportional to the the inverse of
the wire diameter. It is clear from the data shown in
Fig.~\ref{F-Temperature} that nanowires with larger diameters are
shorter than those with smaller diameters, consistent with that
prediction.  We conclude that the nanowires grow via dislocation
climb in the oxide film.

\begin{figure} 
\begin{center}
\includegraphics[width=0.45\textwidth]{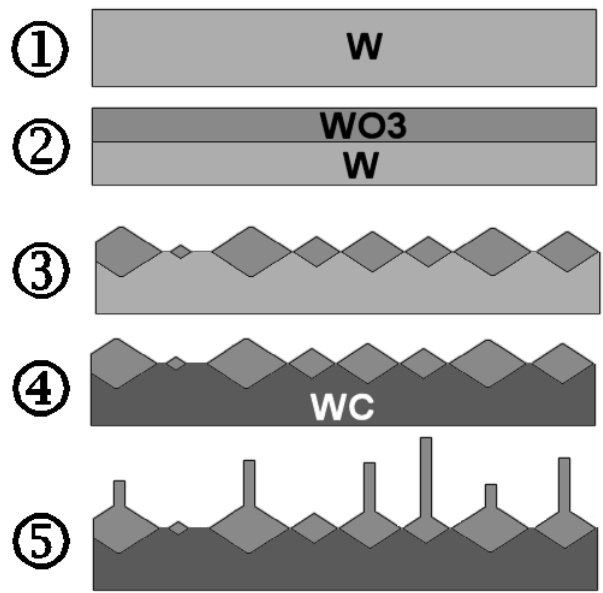}
\caption{\it Proposed growth mechanism: The tungsten substrate (1)
gets oxidized in air (2). In the CVD system at elevated
temperatures the tungsten oxide surface roughens and forms sharp
faceted crystallites with the help of hydrogen (3). Introduced
CH$_{4}$ carburizes the underlaying tungsten. Carbide formation in
turn induces strain at the WC/WO$_{3}$ interface (4). The strain
is relieved by formation of stress-free nanowires (5).}
\label{F-Mechanism}
\end{center}
\end{figure}

While the growth \emph{mechanism} is consistent with whisker
growth, the driving force for whisker formation is less clear.  In
many metal whisker systems (e.g. Sn) the oxidation of the whisker
surface plays an important role in stabilizing the growth of
elongated structures with large surface area to volume ratios. In
the present case the surface is already oxidized and the amount of
oxide appears to be conserved.  One possible driving force for
whisker growth in the present case is interfacial strain:  whisker
growth will convert strained oxide at the W/WO$_3$ interface to
unstrained oxide in the nanowire.  The relief of strain may be one
important factor driving whisker formation.

We propose that the strain driving the whisker
formation is enhanced
by carbide formation at the W/WO$_3$ interface.
In this picture carbide
formation increases the interfacial strain (e.g. the W-W separation in WC
increases by \%4), enhancing nanowire growth. A similar
enhancement has been observed in Sn films subjected to external
loading~\cite{FISHER,FRANKS}. The difference is that in the
present case the strain is induced \emph{chemically}, via carbide
formation at the interface. The process is illustrated
schematically in Fig.~\ref{F-Mechanism}.

In order to clarify the role of the carbide at the interface, we
attempted to re-grow nanowires on the same substrate. That is, we
first grew nanowires in the vacuum system on a pristine filament
using the procedure outlined above. The filament was then
sonicated to remove the nanowires, and left in air for several
days. It was then reintroduced into the vacuum system and
subjected to the nanowire growth process for a second time.
However, no nanowires grew during the second processing. The
filament was irreversibly changed in the first processing,
presumably because the carbide we measure using XPS is formed at
the surface. Note that TEM measurements show no evidence for
carbide formation in the nanowires. Subsequent processing does not
produce nanowires because either, (a) no oxide forms on the
carbide surface, or (b) further carbide formation does not change
the strain state of the native oxide. We conclude that carbide
formation at the W/WO$_3$ \emph{interface} is associated with the
high nanowire yield. Hydrogen pre-treatment is also required for
high yield. As the Fig.~\ref{F-CVD}c shows, hydrogen pretreatment
significantly changes the surface morphology, presumably by
enhancing oxide diffusion. H$_{2}$ is known to enhance oxygen
diffusion in semiconductors~\cite{JONES}. In whisker growth, the
whisker diameter is determined by surface morphology. Hydrogen
pretreatment may help in forming the crystalline grains required
for classical whisker growth.

In conclusion, our results show that exposure to both hydrogen and
methane strongly enhances the formation of nanowires on (natively)
oxidized W films during CVD processing. Structural and chemical
analysis indicates that the nanowires are crystalline WO$_{3}$ and
that WC is formed at the surface of the tungsten substrates. The
cross sectional shape and variation in nanowire length with
diameter are consistent with conventional whisker growth. We
conclude that interfacial strain drives the nanowire formation in
a process analogous to that observed in mechanically strained Sn
films. This wire growth mechanism is expected to be general, and
may be of use in enhancing the yield in other nanowire systems.

We thank Bruce Ek for technical support. Christian Klinke
acknowledges financial support from the Alexander-von-Humboldt
Foundation.

\clearpage

\end{document}